# EVALUATION AND EXPLOITATION OF KNOWLEDGE ROBUSTNESS IN KNOWLEDGE-BASED SYSTEMS

**Mathieu Barcikowski\*, Philippe Pernelle\*,
Arnaud Lefebvre\*, Michel Martinez\*, Jean Renaud°**

*\* Laboratoire PRISMA*
*Université Lyon 1*
*17 rue de France, F-69 Villeurbanne cedex*
*mbarciko@bat710.univ-lyon1.fr,*
*{philippe.pernelle, arnaud.lefebvre}@iutb.univ-lyon1.fr,*
*michel.martinez@univ-lyon1.fr*
*°Equipe ERPI, ENSGSI – INPL*
*8 rue Bastien Lepage B.P. 647 54010 Nancy, France*
*Jean.Renaud@ensgsi.inpl-nancy.fr*

Abstract: Industrial knowledge is complex, difficult to formalize and very dynamic in reason of the continuous development of techniques and technologies. The verification of the validity of the knowledge base at the time of its elaboration is not sufficient. To be exploitable, this knowledge must then be able to be used under conditions (slightly) different from the conditions in which it was formalized. So, it becomes vital for the company to permanently evaluate the quality of the industrial knowledge implemented in the system. This evaluation is founded on the concept of robustness of the knowledge formalized by conceptual graphs. The evaluation method is supported by a computerized tool. *Copyright © 2006 IFAC*

## 1. INTRODUCTION

To ensure its perenniality, the industrial company must remain competitive and improve its capacity of innovation. In this context, the capitalization of expert's knowledge and the preservation of firm know-how become a crucial stake. For that, knowledge engineering provides a set of tools and methods which aim at satisfying this requirement of capitalization. In this study, we consider the problem of the intrinsic validity and the suitability of this industrial knowledge at the time of its use. More precisely, is this knowledge exploitable in a context or use case different from those in which they were defined. To answer this question, we introduce the concept of robustness of the technical knowledge.

First we introduce problems about capitalization, use and evolution of knowledge inside companies. The introduction of the concept of robustness applied to knowledge and the definition of various criteria to evaluate this robustness are presented in section 3.

Finally in the last part, we detail a example of calculation and use of some robustness criteria.

## 2. KNOWLEDGE IN COMPAGNIES

Knowledge present within the company is varied and complex. Its classification and even its significance depend on the point of view of the actors. We use a commonly accepted approach which separates explicit knowledge and tacit knowledge. Explicit knowledge is dedicated to a particular problem. It generally relates to heuristics representing expert experience and is often approachable only with difficulty. More, its mode of formalization does not always meets manufacturing needs. Tacit knowledge (Grundstein, 2001) corresponds to manufacturing expert know-how which is more or less easily formalisable. Knowledge can be essentially obtained by two ways. Some methodologies (MASK (Ermine, 2001)...) propose to use company experts as source of knowledge. But it is also possible to extract knowledge from documents resulting from firm activity (Bourigault et al., 1996) (Assidi, 1998) (Biébow et al., 1999).

Therefore, knowledge is modeled from a particular professional point of view on the field of knowledge, at a given time and according to a specific methodology. So, a divergence can appear between the field of knowledge and its model (the knowledge base).

Capitalization and follow-up of knowledge are difficult because of the constant and rapid evolutions of the industrial firm (Renaud, 2004). The changes in technologies, know-how, environment, manpower and product must be taken into account.

This difficulty is worsen by the constraints due to the access, selection and modeling of the knowledge (Grundstein, 1995) crucial for the company. To be able to use without risk the technical knowledge capitalized in the company, it is necessary to evaluate in which extent knowledge resists the evolutions of the field, i.e. the technical, social and economic evolutions. It is also essential to estimate how knowledge, once modeled, resists the imprecision or divergences due to the problems of selection of the crucial knowledge.

To take into account these evolutions and these difficulties of modelling, it is vital to evaluate the robustness of knowledge (character of nonsensibility to context changes) with respect to system dynamics. This evaluation allows to consider the industrial risk associated to the use of this knowledge in engineering systems or other systems supporting product development. The estimate of the confidence one can grant to knowledge base is able to allow a better anticipation and detection of problems relating to the use and update of knowledge.base and thus to reduce maintenance needs. One can also imagine that a better comprehension of the conditions of knowledge robustness can make it possible to maximise the confidence in the use of this knowledge.

## 3. ROBUSTNESS OF KNOWLEDGE

Robustness is a concept used in many fields (system science, automation ...) and which can be adapted to the approach of knowledge by analogy. Intuitively, one can say that the more knowledge is robust, the more it could be used in a different or uncertain context as regards to the initial context in which it was created. More formally, we approach the definition of knowledge robustness according to two aspects:
– The *temporal* aspect allows to take into account the divergence between the knowledge modelled at a given time and the natural progress in the manufacturing domain,
– The *contextual* aspect permits to take into account the deviation between the knowledge base and the actual field because of modelling uncertainties or errors.

### 3.1 Contextual robustness

Let us consider the case of a knowledge base exploited through an inference engine. If one supposes that the quality of the inference engine is sufficient, then the quality of the supplied answers is only dependent on the quality of the knowledge base (Groot et al., 2000). Thus, a knowledge base is said "robust" if it allows to preserve the quality of the supplied answers, whatever the context of use is.

### 3.2 Temporal robustness

One is interested here in the ability of the knowledge base to resist field evolutions. The knowledge field to model within the knowledge base is depend on the evolution of the real world. Evolutions of the technical, social and economic environment periodically question the current knowledge and know-how. Ideally, any modification of the field should result in a modification of the knowledge base by the experts of the field. In the actual case, this dynamics can be difficult to set up because it can be difficult to have access and to detect these changes.

A knowledge base with robust nature permits to limit the effects of the modifications of the knowledge field. It can preserve sufficient quality without maintenance of the knowledge base in spite of industrial field evolutions.

### 3.3 Criteria of robustness

The level of robustness of knowledge is not evaluated in a global way but it is evaluated by a set of robustness criteria. Each one of these criteria makes it possible to describe an aspect of the robustness of knowledge. It is thus necessary to take into account each one of these criteria to have a global vision of the robustness of knowledge. Among these criteria, we propose criteria of robustness related to the activity of update of the knowledge base. These criteria will be clarified in a more concrete way through the example below.

The criteria related to the activity of update of the knowledge base make it possible to observe the stability of the knowledge base in the course of time. They allow observing zones of the knowledge base which are stabilized because of the maturity reached by knowledge which contain. It also makes it possible to observe zones which were not updated and which requires possibly a modification because of evolution of the field. They finally make it possible to put forward the zones of the base of knowledge which requires a monitoring deepened because of the great activity related to these zones.

In the following, the exposed figures and results come from our system EvaTRoK (Evaluation tool of Robustness of Knowledge). The goal of the EvaTRoK system is to provide a decision aid tool making it possible to improve maintenance of a base of knowledge. We use criteria related to the evolution of the base of knowledge but other criteria can be taken into account (Barcikowski et al., 2004).

## 4. EVALUATION AND VISUALIZATION OK KNOWLEDGE ROBUSTNESS

We will describe how are evaluated the criteria related to the evolution of the knowledge base starting from a concrete example. The knowledge base is described by a conceptual graph. It contains knowledge describing a person thinking of a picture, picture showing a fisherman and a friend on a lake. To simulate the evolution of the knowledge base, a second version of this one is created, bringing a second character thinking of the picture, which is in fact the fisherman painted on the picture. And some additional knowledge relating to the description of the painted scene is added to the description of the picture.

### 4.1 Conceptual graphs

Knowledge will be modelled using conceptual graphs (Sowa, 1984). The model of conceptual graphs is a model of knowledge representation based on the existential graphs of C PEIRCE (Peirce, 1933) and the semantic networks of artificial intelligence. A conceptual graph is a graph composed of two types of nodes, respectively the concept type node and the relation type node.

The model of the conceptual graphs is defined formally by means of an abstract syntax which allows the representation of the graphs according to various notations. The Display Form makes it possible to the users to understand and modify the conceptual graphs more easily than with a representation in the form of logical formulas for example. It is also possible to represent conceptual graphs by using XML such as the editor of conceptual graphs CharGer uses it (Delugach, 2001).

The Fig. 1 represents the sentence "John takes the bus in direction of Boston" by a conceptual graph in its graphical form.

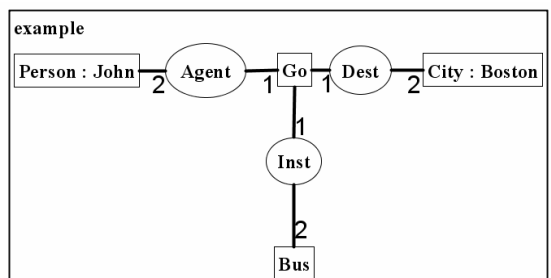

Fig. 1. Example of conceptual graph.

### 4.2 Difference between two conceptual graphs

We seek here to evaluate criteria of robustness which depend to the evolution of the knowledge base. For that we need information relating to the update of the knowledge base. This information can be obtained directly by the use of log file of events which the knowledge base editor generates. If log file is not available, it is possible to obtain this information by calculating the difference between two versions of the base of knowledge. We start from this second assumption, more constraining, for the continuation of the article.

We defined three types of handling which can be carried out on the basis of knowledge. The *addition* type corresponds to the addition of an element in the knowledge base between the current version and the preceding version. The *suppression* type indicates that an element present in the preceding version of the knowledge base is not present any more in the new version. Finally the *modification* type indicates that an element present in the preceding version is always present in the new version of the knowledge base, but also states that the description of this element was modified between the two versions of the base.

To carry out the difference between the two versions of the knowledge base, we assume the assumption that each element of the knowledge base is identified by a single identifier generated by the system.

Thus to list the additions carried out on the knowledge base, it is enough to compare the identifiers of elements present in the new version of the knowledge base and which were not present in the old version. In a similar way, the discovery of the removed elements is carried out by the comparison of the identifiers present in the old version of the knowledge base and which disappeared from the new version of the knowledge base. Finally for the modified elements, their identifiers are at the same time present in the old one and the new version of the knowledge base. To determine if they were modified, it is necessary then to determine if information describing the element was changed. It is also necessary to take into account only relevant information from point of view of knowledge. For example, it is desirable not to take into account information related to the presentation of the element in a graphic interface.

The example of the Fig. 2 shows the result of the calculation of the difference between two versions of a knowledge base. The *Target* part indicates if the update intervenes on an element or a link connecting two elements of the knowledge base. The *Type* part indicates the type of update carried out on the element.

```
Target : SELF / Type : MOD /
From :Person / To : Person
Target : SELF / Type : MOD /
From :Person / To : Person :
John
Target : SELF / Type : ADD /
From :null / To : Person :
Peter
Target : SELF / Type : ADD /
From :null / To : Think
Target : SELF / Type : ADD /
From :null / To : Truit
Target : LINK / Type : ADD /
From :null / To : [contain ->
Lake] : 1
Target : LINK / Type : ADD /
From :null / To : [contain ->
Truit] : 2
```

Fig. 2. Extract of the result of a difference between two versions of a conceptual graph.

The *From* part and the *To* part describe in a textual way the element in the old one and the new version of the knowledge base. These two parts indicate null respectively if the element does not exist or no longer exists. If one interprets the first line of the Fig. 3, one can see that an element suffers a modification, this element being described by the identifier "person".

```
□ e robustness
  □ e criteria
    ⓐ id                cn10
    ⓐ nodetype          concept
    □ e c4
      ⓐ add             0.0
      ⓐ del             0.0
      ⓐ mod             0.0
  □ e criteria
    ⓐ id                cn13
    ⓐ nodetype          concept
    □ e c4
      ⓐ add             0.0
      ⓐ del             0.0
      ⓐ mod             1.0
  □ e criteria
    ⓐ id                cn14
    ⓐ nodetype          concept
    ⊞ e c4
  □ e criteria
    ⓐ id                cn15
    ⓐ nodetype          concept
    □ e c4
      ⓐ add             0.0
      ⓐ del             0.0
      ⓐ mod             1.0
```

Fig. 3. Extract of the result of the evaluation of the criteria of robustness related to the evolution of the knowledge base.

*4.3 Evaluation of criteria of robustness which depend to the evolution of the base of knowledge.*

The criteria related to the activity of update of the knowledge base make it possible to observe which is the stability of the knowledge base in the course of time.
They allow observing zones of knowledge base which are stabilized because of the maturity reached by knowledge they contain. It also makes it possible to highlight zones which were not updated and which requires possibly a modification because of evolution of the field. They finally make it possible to put to locate zones of the base of knowledge which requires deepened monitoring because of the great activity related to these zones.

These criteria are placed in temporal dimension. Indeed, they make it possible to evaluate the quantity of update operated in the course of time, and thus indicate the temporal evolution of the knowledge base.

The example of the Fig. 3 shows us the result of the evaluation of the criteria of robustness related to the evolution of the knowledge base. Each element of the conceptual graph identifiable by an identifier is assigned different values for the criteria of addition, suppression and modification. These values are calculated directly from the result of the difference between the two versions of the knowledge base, i.e. for each element, we carry out the sum of the number of action of additions, modification and suppression applied to this element. For the complex elements such as overlapping graphs or for the whole conceptual graph, the values of their criteria of robustness are obtained by carrying out the sum of the values of the criteria of robustness of the elements which they contain.

Thus in the example the concepts identified cn13 and cn15 have undergone one action of modification in total. They thus do not have undergoes any action of addition or suppression. And the other concepts do not have undergoes any action of update.

*4.4 Exploitation of the results.*

The tool of evaluation and visualization of the robustness aims at providing a decision-aid tool. Used at the time of the stage of maintenance of the knowledge base, it provides a set of means allowing displaying the result of the evaluation of the robustness.

The spatial displaying of conceptual graphs is carried out in two dimensions by means of rectangle, ellipses... representing the concepts, the relations... of the knowledge base and lines representing the links between the elements of the knowledge base.

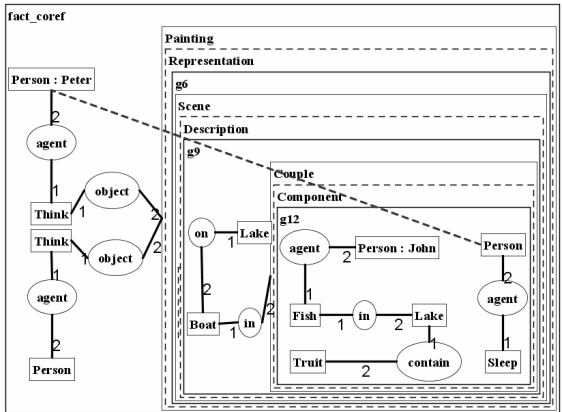

Fig. 4. Spatial view of the knowledge base.

The values of the criteria of robustness are visualized by means of a code colour used to draw each nodes of graph and representing the rate of robustness evaluated for the nodes of the graph, according to the criterion of robustness selected by the expert in charge of the maintenance of the base.

The values of the criteria of robustness are visualized by means of a code colour used to draw each nodes of graph and representing the rate of robustness evaluated for the nodes of the graph, according to the criterion of robustness selected by the expert in charge of the maintenance of the base

The Fig. 4 represents the conceptual graph without additional display information on the evaluation the robustness of the knowledge base. This first view shows the new version of the knowledge base. The user can then choose to display the value of the various calculated criteria of robustness. Within the framework of the criteria presented in this article, one can consequently choose to display the criterion of addition, suppression or modification.

The selection of a criterion causes the colouring of the conceptual graph in order to bring information additional to the expert user. Thus on the fig. 5 and 6, the elements of the conceptual graph are coloured in such a way that the elements which have undergo the greater number of update are seen displayed in a clearer colour.

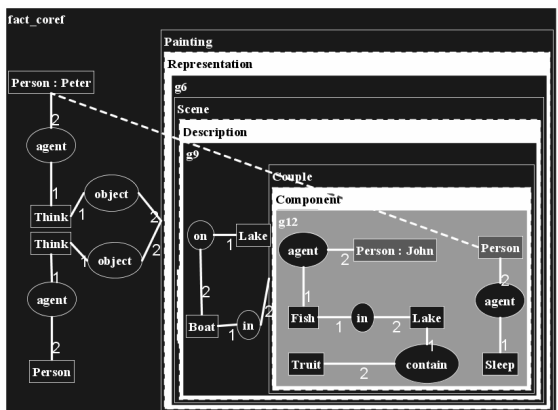

Fig. 5. Augmented spatial view of the knowledge base - modification criterion

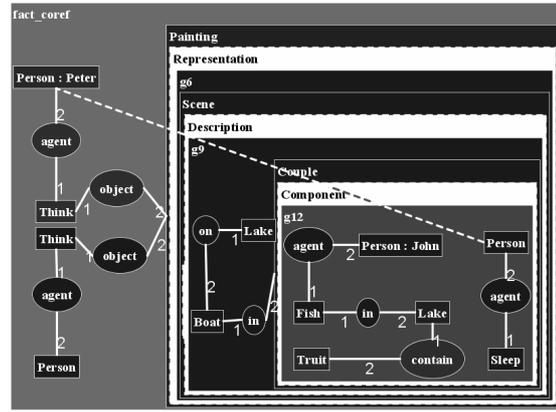

Fig. 6. Augmented spatial view of the knowledge base - addition criterion.

Coloured displaying therefore makes it possible to reveal the least stable zones of the knowledge base. Thus one notes in the Fig. 5 that only the area corresponding to the graph describing the picture was modified. One notes in the Fig. 6 that in the graph corresponding to the description of the scene and the picture of the additions were carried out. From a global point of view, one deduces that the graph describing the picture is not stable. This instability expresses a lack of robustness on the level of this element of knowledge.

## 5. CONCLUSION

For manufacturers, the constitution of company memory represents an essential stage today to develop activities of innovation and creation of new products. But in the industry field, the capture, validation and formalization of knowledge are not enough to ensure that knowledge remains perennial and exploitable at the time of its use. Modelled knowledge must possess a certain degree of robustness which varies in economic time according to technological and human factors.

In this study we introduce two aspects of the robustness of knowledge, i.e. a temporal and a contextual point of view. These two points of view take into account the use and evolution of the knowledge base.

We illustrate the use of robustness criteria related to the evolution of the knowledge base. We describe how to gather information information from traces of the evolution of knowledge base, and then how to interpret this information for the calculation of robustness criteria. Finally we introduce the EvaTRoK system which analyses the knowledge base and achieves an evaluation of robustness criteria. This software aims to help in maintenance of the knowledge, like a decision aid tool.

Our next step will be to propose a complete decision aid tool with a full set of robustness criteria and to test this tool in a in a more important and realistic use case.
.